\documentclass{article}
\begin{document}
\begin{center}
GRAVITATIONAL COUPLINGS FOR GOp-PLANES AND y-Op-PLANES \\ [.25in]
by Juan F. Ospina G.
\end{center}
\begin{center}
ABSTRACT \\ [.25in]
The Wess-Zumino actions for generalized orientifold planes (GOp-planes) and y-deformed orientifold planes (yOp-planes) are presented and two series power expantions are realized from whiches processes that involves GOp-planes,yOp-planes, RR-forms , gravitons and gaugeons , are obtained. Finally non-standard GOp-planes and y-Op-planes are showed.
\end{center}

\section{Introduction}
The results that this paper presents are about gravitational couplings for generalized orientifold planes (GOp-planes) and y-deformed orientifold planes (yOp-planes). The usual orientifold planes do not have gauge fields on their worldvolumes and  . The generalized orientifold planes that this paper consider have SO(2k) Yang-Mills gauge fields-bundles over their corresponding worldvolumes. The aim of the present paper is to display the Wess-Zumino part of the effective action for such generalized orientifold planes.

The usual orientifold planes do not have any kind of topological y-deformation.This paper presents the Wess-Zumino action for  y-deformed orientifold planes.

For the usual orientifold planes the Wess-Zumino action has the following form,which can be derived both from anomaly cancellation arguments and from
direct computation on string scattering amplitudes:   

\begin{center}
{ \mathversion{bold} $ S_{WZ(Op-plane)} = -2^{p-4}\frac{\rm T_p}{\rm kappa}\int_{p+1} C\sqrt{\frac{\rm L(\frac{\rm R_T}{\rm 4})}{\rm L(\frac{\rm R_N}{\rm 4})}}$ }
\end{center}

Where the Mukai vector of RR charges for the usual orientifold p-plane is given by:

\begin{center}
{ \mathversion{bold} $ Q(\frac{\rm R_T}{\rm 4},\frac{\rm R_N}{\rm 4}) =\sqrt{\frac{\rm L(\frac{\rm R_T}{\rm 4})}{\rm L(\frac{\rm R_N}{\rm 4})}}  $ }
\end{center}
In this formula C is the vector of the RR potential forms. L is the Hirzebruch genus that generates the Hirzebruch polynomials which are given in
terms of Pontryaguin classes for real bundles. The Pontryaguin classes are given in terms of the 2-form curvature of the corresponding real bundle. The
formula for Q involves two real bundles over the worldvolume of the usual orientifold plane.  These two bundles are the tangent bundle for the worldvolume and the normal bundle by respect to space-time for such worldvolume. Q is given then in terms of the curvatures for the tangent and
normal bundles and does not have contributions  from the others real bundles
such as SO(2k) Yang-Mills gauge bundles and does not have any kind of topological deformation.

In this paper is presented the Mukay vector of RR charges for a generalized
orientifold planes which have two SO(2k) Yang-Mills gauge bundles on their worldvolumes.  Such vector of RR charges is given by the following formula:

\begin{center}
{ \mathversion{bold} $ Q(\frac{\rm R_T}{\rm 2},\frac{\rm R_N}{\rm 2},\frac{\rm R_E}{\rm 2},\frac{\rm R_F}{\rm 2}) =\sqrt{\frac{\rm A(\frac{\rm R_T}{\rm 2})Mayer(\frac{\rm R_E}{\rm 2})}{\rm A(\frac{\rm R_N}{\rm 2})Mayer(\frac{\rm R_F}{\rm 2})}} $ }
\end{center}

For the generalized  orientifold planes the Wess-Zumino action has the following form:

\begin{center}
{ \mathversion{bold} $ S_{WZ(GOp-plane)} = -2^{p-4}\frac{\rm T_p}{\rm kappa}\int_{p+1} C\sqrt{\frac{\rm A(\frac{\rm R_T}{\rm 2})Mayer(\frac{\rm R_E}{\rm 2})}{\rm A(\frac{\rm R_N}{\rm 2})Mayer(\frac{\rm R_F}{\rm 2})}}$ }
\end{center}

The formula for the vector of RR charges corresponding to a generalized orientifold plane involves now four real bundles over the worldvolume: the 
tangent bundle, the normal bundle and two new SO(2k) YM gauge bundles.
When one of these new SO(2k) bundles is the tangent bundle and the other is the normal bundle, one obtain the usual formula for Q corresponding to the usual orientifold planes using the following identity:

\begin{center}
 
{ \mathversion{bold} $A(\frac{\rm R}{\rm 2})Mayer(\frac{\rm R}{\rm 2}) 
 = L(\frac{\rm R}{\rm 4})$}
\end{center} 

Then, one has:

\begin{center}
{ \mathversion{bold} $ Q(\frac{\rm R_T}{\rm 2},\frac{\rm R_N}{\rm 2},\frac{\rm R_T}{\rm 2},\frac{\rm R_N}{\rm 2}) =\sqrt{\frac{\rm A(\frac{\rm R_T}{\rm 2})Mayer(\frac{\rm R_T}{\rm 2})}{\rm A(\frac{\rm R_N}{\rm 2})Mayer(\frac{\rm R_N}{\rm 2})}} $ }
\end{center}

\begin{center}
{ \mathversion{bold} $ Q(\frac{\rm R_T}{\rm 2},\frac{\rm R_N}{\rm 2},\frac{\rm R_T}{\rm 2},\frac{\rm R_N}{\rm 2}) =\sqrt{\frac{\rm L(\frac{\rm R_T}{\rm 4})}{\rm L(\frac{\rm R_N}{\rm 4})}}=Q(\frac{\rm R_T}{\rm 4},\frac{\rm R_N}{\rm 4}) $ }
\end{center}

In these formulas, A denotes the roof-Dirac genus and Mayer denotes the Mayer class for one SO(2k) YM gauge bundle.

In this paper,also, is presented the Mukay vector of RR charges for the y-deformed  orientifold planes which have topological y-deformations on their worldvolumes.  Such vector of RR charges is given by the following formula:

\begin{center}
{ \mathversion{bold} $ Q(\frac{\rm R_T}{\rm 4},\frac{\rm R_N}{\rm 4},y) =\sqrt{\frac{\rm CHI_y(\frac{\rm R_T}{\rm 4}) )}{\rm CHI_y(\frac{\rm R_N}{\rm 4})}} $ }
\end{center}

For the y-deformed  orientifold planes the Wess-Zumino action has the following form:

\begin{center}
{ \mathversion{bold} $ S_{WZ(yOp-plane)} = -2^{p-4}\frac{\rm T_p}{\rm kappa}\int_{p+1} C\sqrt{\frac{\rm CHI_y(\frac{\rm R_T}{\rm 4}) )}{\rm CHI_y(\frac{\rm R_N}{\rm 4})}}$ }
\end{center}
The formula for the vector of RR charges corresponding to a y-deformed orientifold plane involves now two real bundles over the worldvolume: the 
tangent bundle and  the normal bundle, but in this case one has a topological y-deformation over the worldvolume.
When the parameter y of the topological y-deformation is 1, then one obtain the usual formula for Q corresponding to the usual orientifold planes using the following identity:
\begin{center}
 
{ \mathversion{bold} $ 
 CHI_1(R)= L(R)$}
\end{center} 
Then, one has:
\begin{center}
{ \mathversion{bold} $ Q(\frac{\rm R_T}{\rm 4},\frac{\rm R_N}{\rm 4},1) = \sqrt{\frac{\rm CHI_1(\frac{\rm R_T}{\rm 4}) )}{\rm CHI_1(\frac{\rm R_N}{\rm 4})}}$ }
\end{center}

\begin{center}
{ \mathversion{bold} $ Q(\frac{\rm R_T}{\rm 4},\frac{\rm R_N}{\rm 4},1,) =\sqrt{\frac{\rm L(\frac{\rm R_T}{\rm 4})}{\rm L(\frac{\rm R_N}{\rm 4})}}=Q(\frac{\rm R_T}{\rm 4},\frac{\rm R_N}{\rm 4}) $ }
\end{center}
In these formulas, CHI sub y  denotes the chi-y- genus which when y=1 is the Hirzebruch-genus and when y=0, is the Todd genus.

In the following section the Mukay vector of RR charges for a such generalized orientifold p-plane (GOp-plane) and such y-deformed orientifold p-plane (yOp-plane), will be given in terms of the powers of the curvatures for the four real bunldes involved over the worldvolume   for the GOp-planes and in terms of the powers of the curvatures for the two y-deformed bundles in the case of yOp-planes.

 In the third section are presented the elementary processes corresponding to the power expansion for the two Q's. In the final four section some conclutions are presented about other GOp-planes and yOp-planes  and non-BPS GOp-planes and yOp-planes.

\section{The Power Expantion for Q's}

Let E be a SO(2k)-bundle over the worldvolume of a generalized orientifold plane and consider a formal factorisation for the total Pontryaguin classs of the real bundle E, which has the following form:

\begin{center}
{ \mathversion{bold} $ p(E) = \prod_{i=1}^k(1+y_i^2)$ }
\end{center}
The total Pontryaguin classs of the real bundle E,has the following formal sumarisation in terms of the corresponding Pontryaguin classes: 
\begin{center}
{ \mathversion{bold} $ p(E) = \sum_{j=0}^{\infty}p_j(E) $ }
\end{center}
The total Mayer class for the real bundle E has the following formal factorisation:
\begin{center}
{ \mathversion{bold} $ Mayer(E) = \prod_{i=1}^kcosh(\frac{\rm y_i}{\rm 2})$ }
\end{center}

The total Mayer class for the real bundle E has the following formal sumarisation in terms of the Mayer polynomials which are formed from the corresponding Pontryaguin classes :
\begin{center}
{ \mathversion{bold} $ Mayer(E) = \sum_{j=0}^{\infty}Mayer_j(p_1(E),...,p_j(E)) $ }
\end{center}
The Mayer polynomials are given by:
\begin{center}
{ \mathversion{bold} $ Mayer_0(p_0(E)) = Mayer_0(1)=1 $ }
\end{center}
\begin{center}
{ \mathversion{bold} $ Mayer_1(p_1(E)) = \frac{\rm p_1(E)}{\rm 8} $ }
\end{center}
\begin{center}
{ \mathversion{bold} $ Mayer_2(p_1(E),p_2(E)) = \frac{\rm p_1(E)^2+4p_2(E)}{\rm 384} $ }
\end{center}
\begin{center}
{ \mathversion{bold} $ Mayer_3(p_1(E),p_2(E),p_3(E)) = \frac{\rm p_1(E)^3+12p_1(E)p_2(E)+48p_3(E)}{\rm 46080} $ }
\end{center}
The pontryaguin classes of the real bundle E have the following realizations in terms of the powers of the 2-form curvature for such bundle.  For this curvature  the y's are the eigenvalues:
\begin{center}
{ \mathversion{bold} $  p_1(E)=p_1(R_E) =-\frac{\rm 1}{\rm 8pi^2}trR_E^2 $ }
\end{center}
\begin{center}
{ \mathversion{bold} $  p_2(E)=p_2(R_E) =\frac{\rm 1}{\rm 16pi^4}[\frac{\rm 1}{\rm 8}(trR_E^2)^2-\frac{\rm 1}{\rm 4}trR_E^4] $ }
\end{center}
\begin{center}
{ \mathversion{bold} $  p_3(E)=p_3(R_E) =\frac{\rm 1}{\rm 64pi^6}[-\frac{\rm 1}{\rm 48}(trR_E^2)^3-\frac{\rm 1}{\rm 6}trR_E^6+\frac{\rm 1}{\rm 8}trR_E^2trR_E^4] $ }
\end{center}
Using all these expretions one can to obtain the following expantion:
\begin{center}
\setlength{\baselineskip}{30pt} 
{ \mathversion{bold} $  Mayer(\frac{\rm R_E}{\rm 2}) = 1+\frac{\rm p_1(R_E)}{\rm 32}+\frac{\rm p_1(R_E)^2+4p_2(R_E)}{\rm 6144}+\frac{\rm p_1(R_E)^3+12p_1(R_E)p_2(R_E)+48p_3(R_E)}{\rm 2949120}+...$ }
\end{center}
Now one has the following expantions:
\begin{center}
\setlength{\baselineskip}{30pt} 
{ \mathversion{bold} $  A(\frac{\rm R}{\rm 2}) = 1-\frac{\rm p_1(R)}{\rm 96}+\frac{\rm 7p_1(R)^2-4p_2(R)}{\rm 92160}+...$ }
\end{center}
\begin{center}
\setlength{\baselineskip}{30pt} 
{ \mathversion{bold} $  L(\frac{\rm R}{\rm 4}) = 1+\frac{\rm p_1(R)}{\rm 48}+\frac{\rm -p_1(R)^2+7p_2(R)}{\rm 11520}+...$ }
\end{center}
Using these three expantions it is easy to obtain the following identities:

\begin{center}
 
{ \mathversion{bold} $A(\frac{\rm R}{\rm 2})Mayer(\frac{\rm R}{\rm 2}) 
 = L(\frac{\rm R}{\rm 4})$}
\end{center} 
\begin{center}
 
{ \mathversion{bold} $A(R)Mayer(R) 
 = L(\frac{\rm R}{\rm 2})$}
\end{center} 
\begin{center}
 
{ \mathversion{bold} $A(2R)Mayer(2R) 
 = L(R)$}
\end{center}
\begin{center}
{ \mathversion{bold} $A(2^qR)Mayer(2^qR) 
 = L(2^{q-1}R)$}
\end{center} 
\begin{center}
 
{ \mathversion{bold} $[A(R)2^kMayer(R)]_{top form} 
 = L(R)_{top form}$}
\end{center}

With the help from these identities one has that:
\begin{center}
{ \mathversion{bold} $ \sqrt{\frac{\rm A(\frac{\rm R_T}{\rm 2})Mayer(\frac{\rm R_T}{\rm 2})}{\rm A(\frac{\rm R_N}{\rm 2})Mayer(\frac{\rm R_N}{\rm 2})}}
 = \sqrt{\frac{\rm L(\frac{\rm R_T}{\rm 4})}{\rm L(\frac{\rm R_N}{\rm 4})}}$}

\end{center}
Using all these equations it is easy to obtain the following power expantion for Q:
 
\begin{center}
{ \mathversion{bold} $ \sqrt{\frac{\rm A(\frac{\rm R_T}{\rm 2})Mayer(\frac{\rm R_E}{\rm 2})}{\rm A(\frac{\rm R_N}{\rm 2})Mayer(\frac{\rm R_F}{\rm 2})}}
 = 1+{\frac{\rm (4pi^2alfa)^2}{\rm 1536pi^2}}{(trR_T^2- trR_N^2)} -  \frac{\rm(4pi^2alfa)^2 }{\rm 512pi^2}{(trR_E^2- trR_F^2)}+\frac{\rm (4pi^2alfa)^4}{\rm 4718592pi^4}{{(trR_T^2- trR_N^2)^2}}+\frac{\rm (4pi^2alfa)^4}{\rm 2949120pi^4}{(trR_T^4- trR_N^4)}+\frac{\rm (4pi^2alfa)^4}{\rm 524288pi^4}{{(trR_E^2-trR_F^2)^2}}-\frac{\rm(4pi^2alfa)^4 }{\rm 196608pi^4}{(trR_E^4- trR_F^4)}-\frac{\rm (4pi^2alfa)^4}{\rm 786432pi^4}{(trR_T^2- trR_N^2)(trR_E^2- trR_F^2)}$}

\end{center}
When the bundle E is the tangent bundle and the bundle F is the normal bundle one obtain the usual power expantion for Q corresponding to the usual orientifold
plane:
\begin{center}
{ \mathversion{bold} $ \sqrt{\frac{\rm A(\frac{\rm R_T}{\rm 2})Mayer(\frac{\rm R_T}{\rm 2})}{\rm A(\frac{\rm R_N}{\rm 2})Mayer(\frac{\rm R_N}{\rm 2})}}
 = 1+{\frac{\rm (4pi^2alfa)^2}{\rm 1536pi^2}}{(trR_T^2- trR_N^2)} -  \frac{\rm(4pi^2alfa)^2 }{\rm 512pi^2}{(trR_T^2- trR_N^2)}+\frac{\rm (4pi^2alfa)^4}{\rm 4718592pi^4}{{(trR_T^2- trR_N^2)^2}}+\frac{\rm (4pi^2alfa)^4}{\rm 2949120pi^4}{(trR_T^4- trR_N^4)}+\frac{\rm (4pi^2alfa)^4}{\rm 524288pi^4}{{(trR_T^2-trR_N^2)^2}}-\frac{\rm(4pi^2alfa)^4 }{\rm 196608pi^4}{(trR_T^4- trR_N^4)}-\frac{\rm (4pi^2alfa)^4}{\rm 786432pi^4}{(trR_T^2- trR_N^2)(trR_T^2- trR_N^2)}$} 
\end{center}
\begin{center}
{ \mathversion{bold} $ \sqrt{\frac{\rm A(\frac{\rm R_T}{\rm 2})Mayer(\frac{\rm R_T}{\rm 2})}{\rm A(\frac{\rm R_N}{\rm 2})Mayer(\frac{\rm R_N}{\rm 2})}}
 = 1-{\frac{\rm (4pi^2alfa)^2}{\rm 768pi^2}}{(trR_T^2- trR_N^2)}+\frac{\rm(4pi^2alfa)^4 }{\rm 1179648pi^4}{{(trR_T^2- trR_N^2)^2}}-\frac{\rm 7(4pi^2alfa)^4}{\rm 1474560pi^4}{(trR_T^4- trR_N^4)}$}
\end{center}
\begin{center}
{ \mathversion{bold} $ \sqrt{\frac{\rm A(\frac{\rm R_T}{\rm 2})Mayer(\frac{\rm R_T}{\rm 2})}{\rm A(\frac{\rm R_N}{\rm 2})Mayer(\frac{\rm R_N}{\rm 2})}}
 = \sqrt{\frac{\rm L(\frac{\rm R_T}{\rm 4})}{\rm L(\frac{\rm R_N}{\rm 4})}}$}
\end{center}

For the other hand in the case of the y-Op-plane, the total Chern Class for a  complex n-dimensional bundle V over the worldvolume has the following sumarization:
\begin{center}
{ \mathversion{bold} $ c(V) = \sum_{j=0}^{\infty}c_j(T) $ }
\end{center} 
also, the total Chern Class for the such bundle has the following factorization:

\begin{center}
{ \mathversion{bold} $ c(V) = \prod_{i=1}^{n}(1+x_i)$ }
\end{center}
The  CHI-y- genus for the complex bundle V has the following formal factorisation:
\begin{center}
{ \mathversion{bold} $ CHI_y(V) = \prod_{i=1}^n\frac{\rm(1+yexp(-(y+1)x_i))x_i }{\rm 1-exp(-(y+1)x_i)}$ }
\end{center}

The CHI-y- genus for the complex bundle V has the following formal sumarisation in terms of the y-deformed Todd polynomials which are formed from the corresponding Chern classes and from the polynomials on y :
\begin{center}
{ \mathversion{bold} $CHI_y(V)  = \sum_{j=0}^{\infty}T_j(c_1(V),...,c_j(V),y) $ }
\end{center}

The y-Todd  polynomials are given by:
\begin{center}
{ \mathversion{bold} $ T_0(c_0(V),y) =T _0(1,y)=1 $ }
\end{center}
\begin{center}
{ \mathversion{bold} $ T_1(c_1(V),y) = \frac{\rm (1-y)c_1(V)}{\rm 2} $ }
\end{center}
\begin{center}
{ \mathversion{bold} $ T_2(c_1(V),c_2(V),y) = \frac{\rm (y+1)^2c_1(V)^2+(y^2-10y+1)c_2(V)}{\rm 12} $ }
\end{center}
\begin{center}
{ \mathversion{bold} $ T_3(c_1(V),c_2(V),c_3(V),y) = \frac{\rm -(y+1)^2(y-1)c_1(V)c_2(V)+12y(y-1)c_3(V)}{\rm 24} $ }
\end{center}
\begin{center}
{ \mathversion{bold} $ T_4(c_1(V),c_2(V),c_3(V),c_4(V),y) = \frac{\rm (-y^4+474y^2-124y-1-124y^3)c_4(V)+(y^2-58y+1)(y+1)^2c_1(V)c_3(V)+(y+1)^4(3c_2(V)^2+4c_1(V)^2c_2(V)-c_1(V)^4)}{\rm 720} $ }
\end{center}

Now the relations between the Pontryaguin classes and the Chern Classes for the bundle V are given by the following formulas:

\begin{center}
{ \mathversion{bold} $ p_1(V) = -2c_2(V)+c_1(V)^2$ }
\end{center}

\begin{center}
{ \mathversion{bold} $ p_2(V) = 2c_4(V)-2c_3(V)c_1(V)+c_2(V)^2$ }
\end{center}

Using these relations the y-deformed Todd polynomials can be written as follows:

\begin{center}
{ \mathversion{bold} $ T_0(c_0(V),y) =T _0(1,y)=1 $ }
\end{center}
\begin{center}
{ \mathversion{bold} $ T_1(c_1(V),y) = \frac{\rm (1-y)c_1(V)}{\rm 2} $ }
\end{center}
\begin{center}
{ \mathversion{bold} $ T_2(p_1(V),c_2(V),y) = \frac{\rm (y+1)^2p_1(V)+3(y-1)^2c_2(V)}{\rm 12} $ }
\end{center}
\begin{center}
{ \mathversion{bold} $ T_3(c_1(V),c_2(V),c_3(V),y) = \frac{\rm -(y+1)^2(y-1)c_1(V)c_2(V)+12y(y-1)c_3(V)}{\rm 24} $ }
\end{center}
\begin{center}
{ \mathversion{bold} $ T_4(c_1(V),c_3(V),c_4(V),p_1(V),p_2(V),y) = \frac{\rm -15(y^2+14y+1)(y-1)^2c_4(V)+15(y-1)^2(y+1)^2c_1(V)c_3(V)+(y+1)^4(7p_2(V)-p_1(V)^2)}{\rm 720} $ }
\end{center}
When y=1 the y-deformed Todd polynomials are the same Hirzebruch polynomials:
\begin{center}
{ \mathversion{bold} $ T_0(c_0(V),1) =T _0(1,1)=1=L_0 $ }
\end{center}
\begin{center}
{ \mathversion{bold} $ T_1(c_1(V),1) = \frac{\rm (1-1)c_1(V)}{\rm 2}=0 $ }
\end{center}
\begin{center}
{ \mathversion{bold} $ T_2(p_1(V),c_2(V),1) = \frac{\rm (1+1)^2p_1(V)+3(1-1)^2c_2(V)}{\rm 12}=\frac{\rm p_1(V)}{\rm 3}=L_1(p_1(V)) $ }
\end{center}
\begin{center}
{ \mathversion{bold} $ T_3(c_1(V),c_2(V),c_3(V),1) = \frac{\rm -(1+1)^2(1-1)c_1(V)c_2(V)+12(1-1)c_3(V)}{\rm 24}=0 $ }
\end{center}
\begin{center}
{ \mathversion{bold} $ T_4(c_1(V),c_3(V),c_4(V),p_1(V),p_2(V),1) = \frac{\rm -15(1^2+14+1)(1-1)^2c_4(V)+15(1-1)^2(1+1)^2c_1(V)c_3(V)+(1+1)^4(7p_2(V)-p_1(V)^2)}{\rm 720}= \frac{\rm 7p_2(V)-p_1(V)^2}{\rm 45}=L_2(p_1(V),p_2(V))$ }
\end{center}
Using all these expretions one can to obtain the following expantion:
\begin{center}
\setlength{\baselineskip}{30pt} 
{ \mathversion{bold} $  CHI_y(\frac{\rm R_V}{\rm 4}) = 1+\frac{\rm (1-y)c_1(R_V)}{\rm 8}+\frac{\rm (y+1)^2p_1(R_V)+3(y-1)^2c_2(R_V)}{\rm 192}+\frac{\rm -(y+1)^2(y-1)c_1(R_V)c_2(R_V)+12y(y-1)c_3(R_V)}{\rm 1536}+\frac{\rm -15(y^2+14y+1)(y-1)^2c_4(R_V)+15(y-1)^2(y+1)^2c_1(R_V)c_3(R_V)+(y+1)^4(7p_2(R_V)-p_1(R_V)^2)}{\rm 184320}+...$ }
\end{center}
When the first chern class of V is trivial, one obtain, using again the relations between pontryaguin classes and Chern classes, the following result:
\begin{center}
\setlength{\baselineskip}{30pt} 
{ \mathversion{bold} $  CHI_y(\frac{\rm R_V}{\rm 4}) = 1+\frac{\rm (2(y+1)^2-3(y-1)^2)p_1(R_V)}{\rm 384}+\frac{\rm 12y(y-1)c_3(R_V)}{\rm 1536}+\frac{\rm (-60(y^2+14y+1)(y-1)^2+56(y+1)^4)p_2(R_V)-(-15(y^2+14y+1)(y-1)^2+8(y+1)^4)p_1(R_V)^2}{\rm 1474560}+...$ }
\end{center}
Finally using this last expansion and the relations between the Pontryaguin classes and the 2-form curvature, one can to obtain the following development for the Q of the yOp-planes:
\begin{center}
{ \mathversion{bold} $ \sqrt{\frac{\rm CHI_y(\frac{\rm R_T}{\rm 4}) )}{\rm CHI_y(\frac{\rm R_N}{\rm 4})}}
 = 1+{\frac{\rm (y^2-10y+1)(4pi^2alfa)^2}{\rm 6144pi^2}}{(trR_T^2- trR_N^2)}   +\frac{\rm (4pi^2alfa)^3}{\rm 256}{{y(y-1)(c_3(R_T)- c_3(R_N)}}+\frac{\rm (4pi^2alfa)^4}{\rm 75497472pi^4}{{(y^2-10y+1)^2(trR_T^2-trR_N^2)^2}}-\frac{\rm(4pi^2alfa)^4 }{\rm 188743680pi^4}{(-4y^4-496y^3+1896y^2-496y-4)(trR_T^4- trR_N^4)}$}
\end{center}
When y=1, then one obtain the development for the Q of the usual Op-plane:
\begin{center}
{ \mathversion{bold} $ \sqrt{\frac{\rm CHI_1(\frac{\rm R_T}{\rm 4}) )}{\rm CHI_1(\frac{\rm R_N}{\rm 4})}}
 = 1+{\frac{\rm (1^2-10+1)(4pi^2alfa)^2}{\rm 6144pi^2}}{(trR_T^2- trR_N^2)}   +\frac{\rm (4pi^2alfa)^3}{\rm 256}{{1(1-1)(c_3(R_T)- c_3(R_N)}}+\frac{\rm (4pi^2alfa)^4}{\rm 75497472pi^4}{{(1^2-10+1)^2(trR_T^2-trR_N^2)^2}}-\frac{\rm(4pi^2alfa)^4 }{\rm 188743680pi^4}{(-4-496+1896-496-4)(trR_T^4- trR_N^4)}$}
\end{center} 

\begin{center}
{ \mathversion{bold} $ \sqrt{\frac{\rm CHI_1(\frac{\rm R_T}{\rm 4}) )}{\rm CHI_1(\frac{\rm R_N}{\rm 4})}}
 = 1-{\frac{\rm (4pi^2alfa)^2}{\rm 768pi^2}}{(trR_T^2- trR_N^2)}+\frac{\rm(4pi^2alfa)^4 }{\rm 1179648pi^4}{{(trR_T^2- trR_N^2)^2}}-\frac{\rm 7(4pi^2alfa)^4}{\rm 1474560pi^4}{(trR_T^4- trR_N^4)}$}
\end{center} 
\begin{center}
{ \mathversion{bold} $ \sqrt{\frac{\rm CHI_1(\frac{\rm R_T}{\rm 4}) )}{\rm CHI_1(\frac{\rm R_N}{\rm 4})}}
 =\sqrt{\frac{\rm L(\frac{\rm R_T}{\rm 4})}{\rm L(\frac{\rm R_N}{\rm 4})}} $}
\end{center} 
\section{The Elementary Processes}
The WZ action for the usual orientifold p-plane can be writen as a sum of the WZ actions for three elementary processes:

\begin{center}
{ \mathversion{bold} $ S_{WZ(Op-plane)} = \sum_{j=1}^{3}S_{WZ(Op-plane),j} $ }
\end{center}
The WZ actions for the three elementary processes are given by the following 
expretions:
\begin{center}
{ \mathversion{bold} $ S_{WZ(Op-plane),1} = -2^{p-4}\frac{\rm T_p}{\rm kappa}\int_{p+1} C_{p+1}$ }
\end{center} 
\begin{center}
{ \mathversion{bold} $ S_{WZ(Op-plane),2} = -2^{p-4}\frac{\rm T_p}{\rm kappa}\int_{p+1} C_{p-3}[-({\frac{\rm(4pi^2alfa)^2 }{\rm 768pi^2}}{(trR_T^2- trR_N^2)})]$ }
\end{center}
\begin{center}
{ \mathversion{bold} $ S_{WZ(Op-plane),3} = -2^{p-4}\frac{\rm T_p}{\rm kappa}\int_{p+1} C_{p-7}(\frac{\rm (4pi^2alfa)^4}{\rm 1179648pi^4}{{(trR_T^2- trR_N^2)^2}}-\frac{\rm 7(4pi^2alfa)^4}{\rm 1474560pi^4}{(trR_T^4- trR_N^4)})$ }
\end{center}
The first WZ action describes an elementary process for which the usual orientifold p-plane emites one (p+1)-form RR potential.
The second WZ action describes an elementary process for which the usual
Op-plane absorbs two gravitons and emits one (p-3)-form RR potential.
The third WZ action describes an elementary process for which the Op-plane absorbs four gravitons and emits one (p-7)-form RR potential.
 
From the result of the section two, the WZ action for a generalized orientifold p-plane can be writen as a sum of the WZ actions for some elementary processes:
\begin{center}
{ \mathversion{bold} $ S_{WZ(GOp-plane)} = \sum_{j=1}^{6}S_{WZ(GOp-plane),j} $ }
\end{center}

The WZ actions for the six elementary processes are given by the following 
expretions:
\begin{center}
{ \mathversion{bold} $ S_{WZ(GOp-plane),1} = -2^{p-4}\frac{\rm T_p}{\rm kappa}\int_{p+1} C_{p+1}$ }
\end{center} 
\begin{center}
{ \mathversion{bold} $ S_{WZ(GOp-plane),2} = -2^{p-4}\frac{\rm T_p}{\rm kappa}\int_{p+1} C_{p-3}{\frac{\rm (4pi^2alfa)^2 }{\rm 1536pi^2}}{(trR_T^2- trR_N^2)}$ }
\end{center}
\begin{center}
{ \mathversion{bold} $ S_{WZ(GOp-plane),3} = -2^{p-4}\frac{\rm T_p}{\rm kappa}\int_{p+1} C_{p-3}(-  \frac{\rm (4pi^2alfa)^2}{\rm 512pi^2}{(trR_E^2- trR_F^2)})$ }
\end{center}
\begin{center}
{ \mathversion{bold} $ S_{WZ(GOp-plane),4} = -2^{p-4}\frac{\rm T_p}{\rm kappa}\int_{p+1} C_{p-7}(\frac{\rm(4pi^2alfa)^4 }{\rm 4718592pi^4}{{(trR_T^2- trR_N^2)^2}}+\frac{\rm (4pi^2alfa)^4}{\rm 2949120pi^4}{(trR_T^4- trR_N^4)})$ }
\end{center}
\begin{center}
{ \mathversion{bold} $ S_{WZ(GOp-plane),5} = -2^{p-4}\frac{\rm T_p}{\rm kappa}\int_{p+1} C_{p-7}(\frac{\rm (4pi^2alfa)^4 }{\rm 524288pi^4}{{(trR_E^2-trR_F^2)^2}}-\frac{\rm(4pi^2alfa)^4  }{\rm 196608pi^4}{(trR_E^4- trR_F^4)})$ }
\end{center}
\begin{center}
{ \mathversion{bold} $ S_{WZ(GOp-plane),6} = -2^{p-4}\frac{\rm T_p}{\rm kappa}\int_{p+1} C_{p-7}(-\frac{\rm(4pi^2alfa)^4 }{\rm 786432pi^4}{(trR_T^2- trR_N^2)(trR_E^2- trR_F^2)})$ }
\end{center}

The first WZ action describes an elementary process for which the generalized orientifold p-plane emites one (p+1)-form RR potential.
The second WZ action describes an elementary process for which the generalized
Op-plane absorbs two gravitons and emits one (p-3)-form RR potential.
The third WZ actuib describes an elementary process for which the generalized Op-plane absorbs two gaugeons and emits one (p-3)-form RR potential.
The fourth WZ action describes an elementary process for which the GOp-plane absorbs four gravitons and emits one (p-7)-form RR potential. 
The fifth WZ action describes an elementary process for which the GOp-plane absorbs four gaugeons and emits one (p-7)-form RR potential.
The sixth WZ action describes an elementary process for which the GOp-planes absorbs two gravitons and two gaugeons and emits one (p-7)-form RR potential.

When the gaugeons corresponding to the bundles E and F are the same gravitons corresponding to the bundles T and N respectively, then the six elementary process for the GOp-plane are reduced to the usuals three elementary process for the usual Op-plane: Op-plane emites one (p+1)-form RR potential,Op-plane
absorbs two gravitons and emits one (p-3)-form RR potential; and, Op-plane absorbs four gravitons and emits one (p-7)-form RR potential.

Of other hand, from the result of the section two, the WZ action for a y-deformed orientifold p-plane can be writen as a sum of the WZ actions for some elementary processes:
\begin{center}
{ \mathversion{bold} $ S_{WZ(yOp-plane)} = \sum_{j=1}^{4}S_{WZ(yOp-plane),j} $ }
\end{center}
The WZ actions for the four elementary processes are given by the following 
expretions:
\begin{center}
{ \mathversion{bold} $ S_{WZ(yOp-plane),1} = -2^{p-4}\frac{\rm T_p}{\rm kappa}\int_{p+1} C_{p+1}$ }
\end{center} 
\begin{center}
{ \mathversion{bold} $ S_{WZ(yOp-plane),2} = -2^{p-4}\frac{\rm T_p}{\rm kappa}\int_{p+1} C_{p-3}[-({\frac{\rm(y^2-10y+1)(4pi^2alfa)^2 }{\rm 6144pi^2}}{(trR_T^2- trR_N^2)})]$ }
\end{center}
\begin{center}
{ \mathversion{bold} $ S_{WZ(yOp-plane),3} = -2^{p-4}\frac{\rm T_p}{\rm kappa}\int_{p+1} C_{p-5}(\frac{\rm (4pi^2alfa)^3}{\rm 256}{{y(y-1)(c_3(R_T)- c_3(R_N)))}}-$ }
\end{center}
\begin{center}
{ \mathversion{bold} $ S_{WZ(yOp-plane),4} = -2^{p-4}\frac{\rm T_p}{\rm kappa}\int_{p+1} C_{p-7}(\frac{\rm (4pi^2alfa)^4}{\rm 75497472pi^4}{{(y^2-10y+1)^2(trR_T^2-trR_N^2)^2}}-\frac{\rm(4pi^2alfa)^4 }{\rm 188743680pi^4}{(-4y^4-496y^3+1896y^2-496y-4)(trR_T^4- trR_N^4)})$ }
\end{center}

The first WZ action describes an elementary process on which the yOp-plane emites one (p+1)-form RR potential.
The second WZ action describes an elementary process for which the y-deformed
Op-plane absorbs two gravitons and emits one (p-3)-form RR potential.
The third WZ action describes an elementary process for which the y-deformed
Op-plane absorbs three gravitons and emits one (p-5)-form RR potential.
The fourth WZ action describes an elementary process for which the yOp-plane absorbs four gravitons and emits one (p-7)-form RR potential.
When y=1,then the four elementary process for the yOp-plane are reduced to the usuals three elementary process for the usual Op-plane: Op-plane emites one (p+1)-form RR potential,Op-plane
absorbs two gravitons and emits one (p-3)-form RR potential; and, Op-plane absorbs four gravitons and emits one (p-7)-form RR potential.
 
\section{Conclutions}

The WZ action for the GOp-planes can be modified or extended by various ways.
When the bundles haven non-trivial second Stiefel-Whitney classes one can to write the following WZ action which incorporates an effect of the magnetic monopoles:

\begin{center}
{ \mathversion{bold} $ S_{WZ} = -2^{p-4}\frac{\rm T_p}{\rm kappa}\int_{p+1} C\sqrt{\frac{\rm A(\frac{\rm R_T}{\rm 2})Mayer(\frac{\rm R_E}{\rm 2})e^{\frac{\rm d_1}{\rm 2}}}{\rm A(\frac{\rm R_N}{\rm 2})Mayer(\frac{\rm R_F}{\rm 2})e^{\frac{\rm d_2}{\rm 2}}}}$ }
\end{center}

where:

\begin{center}
{ \mathversion{bold} $ d_1 = reduction.mod.2(w_2(T)+w_2(E))$ }
\end{center}

\begin{center}
{ \mathversion{bold} $ d_2 = reduction.mod.2(w_2(N)+w_2(F))$ }
\end{center}

This action describes processes on which the GOp-plane emites RR-forms and absorbs gravitons, gaugeons and magnetic monopoles.

From the other side one can to write the following actions for GOp-planes non 
standard:

\begin{center}
{ \mathversion{bold} $ S_{WZ} = 2^{p-4}\frac{\rm T_p}{\rm kappa}\int_{p+1} C(2\sqrt{\frac{\rm A(R_T)}{\rm A(R_N)}}-\sqrt{\frac{\rm A(\frac{\rm R_T}{\rm 2})Mayer(\frac{\rm R_E}{\rm 2})}{\rm A(\frac{\rm R_N}{\rm 2})Mayer(\frac{\rm R_F}{\rm 2})}})$ }
\end{center}

\begin{center}
{ \mathversion{bold} $ S_{WZ} = \frac{\rm T_p}{\rm kappa}\int_{p+1} C(\sqrt{\frac{\rm A(R_T)}{\rm A(R_N)}}-2^{p-4}\sqrt{\frac{\rm A(\frac{\rm R_T}{\rm 2})Mayer(\frac{\rm R_E}{\rm 2})}{\rm A(\frac{\rm R_N}{\rm 2})Mayer(\frac{\rm R_F}{\rm 2})}})$ }
\end{center}
In the same way, one can to write the following actions for yOp-planes non standard:
\begin{center}
{ \mathversion{bold} $ S_{WZ} = 2^{p-4}\frac{\rm T_p}{\rm kappa}\int_{p+1} C(2\sqrt{\frac{\rm A(R_T)}{\rm A(R_N)}}-\sqrt{\frac{\rm CHI_y(\frac{\rm R_T}{\rm 4}) }{\rm CHI_y(\frac{\rm R_N}{\rm 4})}})$ }
\end{center}
\begin{center}
{ \mathversion{bold} $ S_{WZ} = \frac{\rm T_p}{\rm kappa}\int_{p+1} C(\sqrt{\frac{\rm A(R_T)}{\rm A(R_N)}}-2^{p-4}\sqrt{\frac{\rm CHI_y(\frac{\rm R_T}{\rm 4} )}{\rm CHI_y(\frac{\rm R_N}{\rm 4})}})$ }
\end{center}
These actions correspond respectively to the Sp-type yOp-planes and the yOp-planes that give rise to gauge symmetries of type SO(2n+1).  Such non-standard yOp-planes are building from combinations of the D-p-branes and standard yOp-planes.

By combination of Dp-branes, Op-planes, GOp-planes  and yOp-planes one can to have gauge teories with symmetries Sp and SO-odd whose WZ actions are give respectively by:
\begin{center}
{ \mathversion{bold} $ S_{WZ} = 2^{p-4}\frac{\rm T_p}{\rm kappa}\int_{p+1} C(2\sqrt{\frac{\rm A(R_T)}{\rm A(R_N)}}-\frac{\rm 1}{\rm 3}(\sqrt{\frac{\rm CHI_y(\frac{\rm R_T}{\rm 4}) }{\rm CHI_y(\frac{\rm R_N}{\rm 4})}}+\sqrt{\frac{\rm L(\frac{\rm R_T}{\rm 4}) }{\rm L(\frac{\rm R_N}{\rm 4})}}+\sqrt{\frac{\rm A(\frac{\rm R_T}{\rm 2})Mayer(\frac{\rm R_E}{\rm 2})}{\rm A(\frac{\rm R_N}{\rm 2})Mayer(\frac{\rm R_F}{\rm 2})}}))$ }
\end{center}
\begin{center}
{ \mathversion{bold} $ S_{WZ} = \frac{\rm T_p}{\rm kappa}\int_{p+1} C(\sqrt{\frac{\rm A(R_T)}{\rm A(R_N)}}-2^{p-4}\frac{\rm 1}{\rm 3}(\sqrt{\frac{\rm CHI_y(\frac{\rm R_T}{\rm 4} )}{\rm CHI_y(\frac{\rm R_N}{\rm 4})}}+\sqrt{\frac{\rm L(\frac{\rm R_T}{\rm 4}) }{\rm L(\frac{\rm R_N}{\rm 4})}}+\sqrt{\frac{\rm A(\frac{\rm R_T}{\rm 2})Mayer(\frac{\rm R_E}{\rm 2})}{\rm A(\frac{\rm R_N}{\rm 2})Mayer(\frac{\rm R_F}{\rm 2})}}))$ }
\end{center}

Finally one can to think about non-BPS GOp-planes  and non-BPS yOp-planes with the tachyon effect.

In conclution gauge theories with symmetries SO-even,Sp and SO-odd can be obtained from the combination of the Dp-branes,Op-planes, GOp-planes and yOp-planes of the string theory.

\section{References}

\setlength{\baselineskip}{20pt}
About WZ action for usual orientifold planes:

K. Dasgupta, D. Jatkar and S. Mukhi, Gravitational couplings and Z2 orientifolds, Nucl. Phys. B523 (1998) 465, hep-th/9707224.

K. Dasgupta and S. Mukhi, Anomaly inflow on orientifold planes, J. High Energy Phys. 3 (1998) 4, hep-th/9709219.

J. Morales, C. Scrucca and M. Serone, Anomalous couplings for D-branes and O-planes, hep-th/9812071.

B.Stefanski,Jr., Gravitational Couplings of D-branes and O-planes, hep-th/9812088

Ben Craps and Frederik Roose, (Non-)Anomalous D-brane and O-plane couplings:the normal bundle,  hep-th/9812149.

About WZ action for non-standard orientifold planes:

Sunil Mukhi and Nemani V. Suryanarayana,  Gravitational Couplings, Orientifolds and M-Planes,  hep-th/9907215

About Mayer class , Mayer integrality theorem and CHI-y-genus:

F. Hirzebruch, Topological Methods in Algebraic Geometry, 1978

Christian Bar,  Elliptic Symbols, december 1995, Math. Nachr. 201, 7-35 (1999)

Taras E. Panov,  Calculation of Hirzebruch Genera for manifolds acted on by the Group Z/p via invariantes of the action, math.AT/9909081

\setlength{\baselineskip}{50pt}   
\end{document}